\documentclass[american,10pt, aps, prl, reprint, showpacs]{revtex4-1}
\usepackage[T1]{fontenc}
\usepackage[utf8]{inputenc}
\usepackage{color}
\usepackage{amsmath}
\usepackage{amssymb}
\usepackage{graphicx}
\usepackage{esint}

\makeatletter

\@ifundefined{textcolor}{}
{%
 \definecolor{BLACK}{gray}{0}
 \definecolor{WHITE}{gray}{1}
 \definecolor{RED}{rgb}{1,0,0}
 \definecolor{GREEN}{rgb}{0,1,0}
 \definecolor{BLUE}{rgb}{0,0,1}
 \definecolor{CYAN}{cmyk}{1,0,0,0}
 \definecolor{MAGENTA}{cmyk}{0,1,0,0}
 \definecolor{YELLOW}{cmyk}{0,0,1,0}
}

\makeatother

\usepackage{babel}
\begin{document}

\title{Optical vortices: the concept of topological potential and analogies
with two-dimensional electrostatics}

\author{Anderson M. Amaral}

\author{Edilson L. Falcão-Filho}

\author{Cid B. de Araújo}

\affiliation{Departamento de Física, Universidade Federal de Pernambuco, 50670-901
Recife, PE, Brazil}

\email{Corresponding author: cid@df.ufpe.br}

\begin{abstract}
We show how the phase profile of a distribution of topological charges
(TC) of an optical vortex (OV) can be described by a potential analogous
to the Coulomb's potential for a distribution of electric charges
in two-dimensional electrostatics. From what we call the Topological
Potential (TP), the properties of TC multipoles and a 2D radial distribution
were analyzed. The TC multipoles have a transverse profile that is
topologically stable under propagation and may be exploited in optical
communications; on the other hand, the 2D distributions can be used
to tune the transverse forces in optical tweezers. Considering the
analogies with the electrostatics formalism, it is also expected that
the TP allows the tailoring of OV for specific applications.
\end{abstract}


\maketitle
Vortices are ubiquitous in nature, and may appear when a physical
field rotates around an axis. They play an important role in fluid
turbulence \cite{ShatsTurbulenceinFluids2014}, atmospheric phenomena
\cite{DritschelVortexinStratosphere2006}, superconductors/superfluids
\cite{abrikosovnobel2004,Lorenz_Complex_Ginzburg_Landau2002} and
optics \cite{Padgett2011}. Indeed, optical vortices (OV) are unique
because the electric field of light evolves linearly in free-space,
and therefore, OV form a prototype system for understanding the general
properties of vortices \cite{Padgett_IsolatedOV2010}. Since the seminal
work by Allen \emph{et al.} \cite{allenorbital1992} OV are subject
of intense research, being applied in areas as diverse as optical
tweezers, laser traps and atom guides \cite{DholakiaShapingOpticalManipulation2011,Pruvost2010,DholakiaAtomGuiding2006},
to excite surface plasmons \cite{brasselettopological2013}, and in
classical and quantum communications \cite{WillnerTerabitFreeSpace2012,WalbornAlignment-FreeOAMQuantumComm2012}.

A quantity present in OV is the topological charge (TC), which measures
the number of twists the electromagnetic field gives around a given
axis in a wavelength. The TC is a non-zero integer for a vortex, and
characterizes their non-trivial topology \cite{Nakahara}. However,
while the TC determines the vortex homotopy class, it does not determine
the geometry; OV can possess the same TC but distinct geometries \cite{Shaping_Optical_Beams}. 

In this work we demonstrate that a distribution of TC in OV can be
described in a similar way to that of a two-dimensional (2D) distribution
of electric charges. The formalism herein presented may be used to
design and to understand paraxial OV in unusual geometries and also
should give insights for vortex properties in other physical systems.

Cylindrical coordinates, where $\mathbf{r}=\mathbf{r}\left(r,\phi,z\right)$,
are considered throughout the paper. In OV, the Poynting vector (and
also the beam wavefront) rotates around the propagation axis $r=0$.
The phase of the electric field $\mathbf{E}\left(\mathbf{r}\right)$
is not defined at $r=0$, so $\mathbf{E}\left(r=0\right)=0$. When
an optical vortex is an eigenstate of the light orbital angular momentum
(OAM), the TC is equal to the OAM per photon \cite{Characterization_of_TC_and_OAM_in_SOV}.
A point TC of charge $q$ is associated with a cylindrical vortex
profile, and appears in $\mathbf{E}\left(\mathbf{r}\right)$ as a
phase term $e^{iq\phi}$, where $\phi$ is the azimuthal angle. The
analogy between an electrostatic charge distribution and a TC arrangement
is considered here by noticing that adding two TCs $q_{1}$ and $q_{2}$
at the same point originates a phase $e^{i\left(q_{1}+q_{2}\right)\phi}$.
So, the composition of TC is additive in the corresponding electric
field. Also, since the total TC is usually conserved in physical processes,
OV are resilient to atmospheric turbulence \cite{WangAtmosphericTurbulence}
and decoherence \cite{pugatchtopological2007}. Analogously, the effect
of two electric charges placed at a point is also additive, and the
total electric charge is a conserved quantity.

The azimuthal phase term constrains the beam intensity profile near
$r=0$ at $z=0$. We assume that the vortex is inside a smooth envelope
$A\left(\mathbf{r}\right)$ around the origin such that $\left.\vec{\nabla}_{\perp}A\left(\mathbf{r}\right)\right|_{r=0}\approx\mathbf{0}$,
where $\vec{\nabla}_{\perp}$ is the gradient taken over the plane
$r\phi$. The electric field $\mathbf{E}\left(\mathbf{r}\right)$
for a monochromatic and homogeneous linearly polarized light beam
propagating along $\hat{z}$ axis on the paraxial regime is represented
by 
\begin{equation}
\mathbf{E}\left(\mathbf{r}\right)=\hat{x}\mathcal{E}\left(\mathbf{r}\right)e^{-i\left(kz-\omega t\right)}=\hat{x}f\left(r\right)e^{iq\phi}A\left(\mathbf{r}\right)e^{-i\left(kz-\omega t\right)},\label{eq:Field}
\end{equation}
and must satisfy the slowly varying envelope approximation (SVEA)
\begin{equation}
2ik\frac{\partial}{\partial z}\mathcal{E}\left(\mathbf{r}\right)=\nabla_{\perp}^{2}\mathcal{E}\left(\mathbf{r}\right).\label{eq:SVEA}
\end{equation}

Equation \eqref{eq:SVEA} implies that for any $q$ at $z=0$, where
$\frac{\partial}{\partial z}\mathcal{E}\left(\mathbf{r}\right)=0$,
$\nabla_{\perp}^{2}\left(fe^{iq\phi}\right)=0.$ So, $f=r^{\left|q\right|}$
at $r\approx0$. For $q>0$ , one may verify that the vortex term
$r^{q}e^{iq\phi}$ is the polar representation of the complex number
$u^{q}$, while for $q<0,$ $r^{\left|q\right|}e^{-i\left|q\right|\phi}=\left(u^{*}\right){}^{\left|q\right|}$.
Therefore, we assume that a TC of $q>0$ can be represented by $u^{q}$.
However, since the topological properties are invariant under small
continuous geometric perturbations, $u^{q}$ and $\left(u-\delta u\right)^{q}$
will roughly represent the same vortex for a sufficiently small $\delta u$.
Also, if $q=q_{1}+q_{2}$, we may consider that $u^{q}\approx\left(u-\delta u_{1}\right)^{q_{1}}\left(u-\delta u_{2}\right)^{q_{2}}$
for small $\delta u_{i}$ ($i=1,2$) \cite{Shaping_Optical_Beams}.
A possible generalization of these transformations at $z=0$ is given
by 
\begin{equation}
\exp V\left(r,\phi,0\right)=\prod_{i=1}^{N}\left(u-u_{i}\right)^{q/N}=R\exp\left(i\Phi_{T}\right).
\end{equation}

In the limit $q\rightarrow0,\, N\rightarrow\infty$ with $q/N$ constant,
it is obtained that $V\left(r,\phi\right)=\int da'\rho\left(u'\right)\log\left(u-u'\right)$,
where $\rho\left(u'\right)$ is an effective TC density per unit area
at $u'=r'e^{i\phi'}$. This expression for $V$ is formally identical
to the 2D electrostatic potential \cite{Panofsky,Smythe}. By extending
the previous calculation to include $q<0$, we have the more general
expression, 
\begin{equation}
V\left(r,\phi,0\right)=\int da'\left[\left|\rho\left(u'\right)\right|\log\left|u-u'\right|+i\rho\left(u'\right)\arg\left(u-u'\right)\right].\label{eq:Topological potential}
\end{equation}

Considering the similarity with the 2D electrostatics Coulomb potential,
we interpret $V\left(r,\phi,0\right)$ as a Topological Potential
(TP) due to an arbitrary TC distribution over the plane $z=0$. Accordingly
to this interpretation, the TP of a TC distribution on a plane is
related to the potential associated with infinite linear distributions
of electric charges. Of course there are some important differences.
For example, the fact that $R=\exp\left\{ \mathfrak{Re}\left[V\left(\mathbf{r}\right)\right]\right\} $
must be finite everywhere implies that $\mathfrak{Re}\left[V\left(\mathbf{r}\right)\right]$
must be always positive and cannot include the TC signal. In other
terms, the intensity profile is not sufficient to determine the topological
properties of an optical vortex. However, the phase profile, $\mathfrak{Im}\left[V\left(\mathbf{r}\right)\right]$,
contains all the topological properties, and it also constrains the
intensity profile. It must be remarked that although $\mathfrak{Re}\left[V\left(\mathbf{r}\right)\right]$
is only an approximation valid near $r=0$, $\mathfrak{Im}\left[V\left(\mathbf{r}\right)\right]$
is exact. Another important difference between the TP and electrostatics
is that the TC is necessarily a discrete quantity and, as will be
shown later, the continuum generalization implied by Eq. \eqref{eq:Topological potential}
indicates that in the general case $\rho\left(u'\right)$ is not a
direct map of the TC distribution. As a remark to our interpretation
of Eq. \eqref{eq:Topological potential}, the reader should notice
that it was insinuated in \cite{berrydislocations1974} that the
vortex phase could be understood as a potential, and the TP introduced
in this work extends this concept by adding the spatial structure.

Another important reason to consider Eq. \eqref{eq:Topological potential}
as a potential comes from the expression for the transverse paraxial
momentum density of light, $\mathbf{p}_{\perp}$. It is possible to
write in our notation that \cite{allenorbital1992} 
\begin{equation}
\mathbf{p}_{\perp}\propto\mathfrak{Im}\left\{ \boldsymbol{\nabla}_{\perp}V\left(\mathbf{r}\right)\right\} I\left(\mathbf{r}\right).\label{eq:Transverse momentum}
\end{equation}

In a semiclassical interpretation of Eq. \eqref{eq:Transverse momentum},
we may understand the intensity profile $I\left(\mathbf{r}\right)$
as the probability of finding a photon at a position $\mathbf{r}$,
while $\mathfrak{Im}\left\{ \boldsymbol{\nabla}_{\perp}V\left(\mathbf{r}\right)\right\} $
represents the local transverse momentum of the photon. Thus, considering
that such light beam can transfer this momentum to an object, the
associated force term would be proportional to the gradient of the
TP.

An application for Eq. \eqref{eq:Topological potential} is to shape
the core of an optical vortex at $z=0$. Assuming that at $r\sim0$
the vortex envelope is Gaussian \textcolor{black}{}%
\footnote{\textcolor{black}{Even our experimental beam having a flat-top amplitude
profile, the Gaussian corresponds to the first correction to the condition
$\vec{\nabla}_{\perp}A\approx\mathbf{0}$ near $r\sim0$. Explicitly,
$A=1-r^{2}/2\approx e^{-r^{2}/2}$.}%
}, $A\left(\mathbf{r}\right)\sim e^{-r^{2}/2}$, the dark core profile
of the vortex is described by $\hat{r}\cdot\vec{\nabla}_{\perp}\mathcal{E}\left(\mathbf{r}\right)=0$,
\textcolor{black}{which by using the Cauchy-Riemann conditions gives}
\begin{equation}
r_{\mbox{core}}^{2}\left(\phi\right)=\left.\frac{\partial}{\partial\phi}\int da'\left|\rho\left(u'\right)\right|\arg\left(u-u'\right)\right|_{r=r_{core}\left(\phi\right)}.\label{eq:vortex_shape}
\end{equation}

This sort of relation between the phase profile and vortex core geometry
was empirically inferred from experimental data \cite{Grier2003}
and verified under more general conditions in \cite{NiuDeterministicOVShapingViaLWD2005}.
However, in the present formalism it arises naturally. Since we did
not consider aperture effects, Eq. \eqref{eq:vortex_shape} is expected
to be valid when $\rho$ is concentrated near $r\approx0$. If $\rho\left(\mathbf{r}\right)$
has the same sign for all $\mathbf{r}$, Eq. \eqref{eq:vortex_shape}
simplifies to $r_{\mbox{core}}^{2}=\left|\partial\Phi_{T}/\partial\phi\right|$,
where the azimuthal derivative is called Local Circulation (LC) and
has an important meaning. It is locally proportional to the classical
OAM density and for a point TC it gives exactly the total OAM per
photon (TC) $q$ \cite{Characterization_of_TC_and_OAM_in_SOV}. The
LC in the present context is the local OAM as discussed in \cite{Characterization_of_TC_and_OAM_in_SOV}.
The vortex radius may be adjusted by tuning the LC to obtain OV with
designed shapes \cite{Grier2003,NiuDeterministicOVShapingViaLWD2005}.
It is also possible, by distributing the TCs over given geometrical
patterns to obtain more general OV profiles as lines, corners and
triangles \cite{Shaping_Optical_Beams}. These procedures can shape
the OV profile at $z=0$. However, the necessary conditions for beam
profile stability under propagation for beams produced via a TCs distribution
remains as an open question. The connections between the TP and the
theory of spiral light beams \cite{AbramochkinSpiralLightBeams2004}
may provide an answer to this point. But this discussion is outside
the scope of the present paper.

An electrostatics-related feature from Eq. \eqref{eq:Topological potential}
is the existence of TCs multipoles. Since multipoles are usually composed
of oppositely charged TCs, they may annihilate under propagation and
may be unstable \cite{WoldDipoleOnFocus1967,rouxdynamical1995,rouxspatial2004}.
However, by expanding the integration kernels of Eq. \eqref{eq:Topological potential}
in terms of circular harmonics, one may verify that a vortex may be
represented in general as $V\left(r,\phi,0\right)=A_{0}\left(r\right)+iB_{0}\left(r,\phi\right)+\sum_{j=1}^{\infty}\left[A_{j}\left(r\right)\cos\left(j\phi\right)+B_{j}\left(r\right)\sin\left(j\phi\right)\right]$,
where $A_{j}\left(r\right),\, B_{j}\left(r\right)$ are complex functions.
We remark that point multipoles with $A_{j}$ and $B_{j}$ proportional
to $r^{-\left(j+1\right)}$ are not meaningful for OV because these
terms imply that when $r\rightarrow0$, both $\mathfrak{Re}\left[V\left(\mathbf{r}\right)\right]$
and $\mathfrak{Im}\left[V\left(\mathbf{r}\right)\right]$ diverge.
We consider that a pure TC multipole is represented by 
\begin{equation}
\Phi_{T}=\alpha\sin\left(j\phi+\beta\right)/j,\label{eq:mutipole-phase-profile}
\end{equation}
where $\alpha$ is constant and $j\neq0$. For an azimuthally periodic
solution of Eq. \eqref{eq:Field} $j$ must be an integer. A fractional
$j$ leads to a line of phase discontinuity similar to those of \cite{Berry2004,Padgettfractionavortex2004}.
$\beta$ is an orientation offset while $\alpha$ determines vortex
core local radius via Eq. \eqref{eq:vortex_shape}, such that 
\begin{equation}
r_{\mbox{core}}\left(\phi\right)=\sqrt{\left|\alpha\cos\left(j\phi+\beta\right)\right|}.
\end{equation}

To produce and characterize some experimental consequences of the
previous discussion, we modulated the wavefront of a 800 nm fiber-coupled
laser diode by using a liquid crystal phase-only spatial light modulator
(SLM) in one arm of a Michelson interferometer, with the same experimental
setup and detection scheme as previously described in \cite{Characterization_of_TC_and_OAM_in_SOV}.
 Unless otherwise stated, the data was collected at the SLM image
plane ($z=0$ cm). The phase profiles on the SLM were composed of
a carrier wave, a circular apodization with a fixed radius and the
phase of interest.

\begin{figure}[h]
\begin{centering}
\includegraphics[width=8cm]{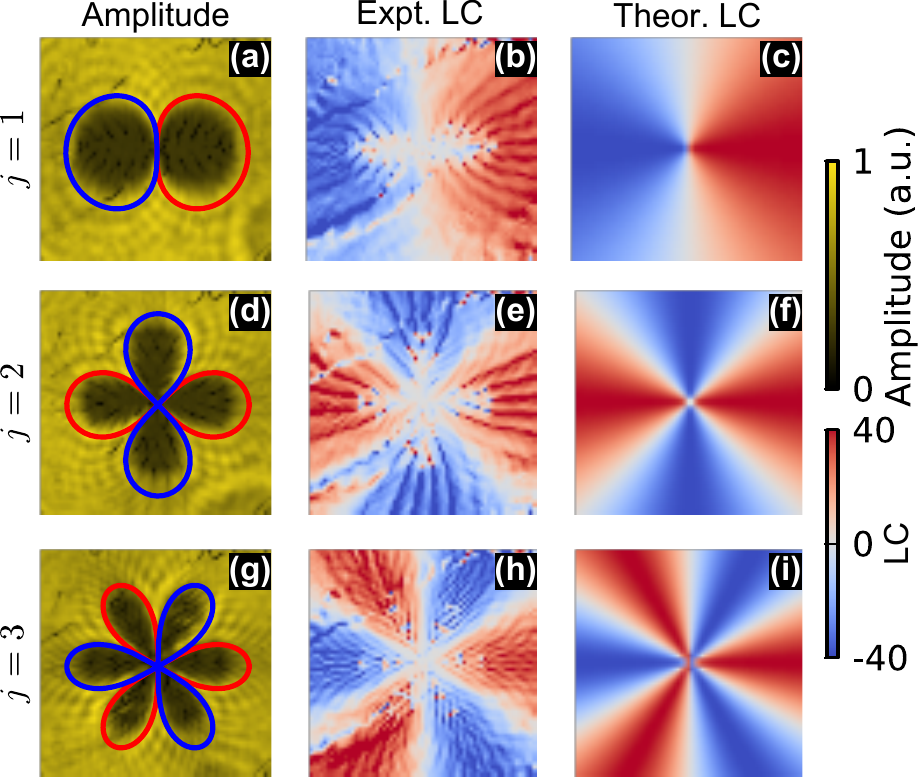}
\par\end{centering}

\protect\caption{Multipoles of TC at $z=0$ cm. The rows correspond to the data for
a dipole (a, b, c), quadrupole (d, e, f) and an hexapole (g, h, i).
In columns we display the beam amplitude profiles (a,d,g), the experimental
LC (b, e, h) and theoretical LC obtained from Eq. \eqref{eq:mutipole-phase-profile}
(c, f, i). The solid lines in (a, d, g) corresponds to the expected
OV core profile from Eq. \eqref{eq:vortex_shape}, and their colors
(red, blue) represent the enclosed TC sign (+, -). \label{fig:multipoles in z0}}
\end{figure}

We show in Figs. \ref{fig:multipoles in z0} (a, d, g) the experimental
intensity profiles for TC multipoles of order $j=$1, 2 and 3 by applying
the phase profile of Eq. \eqref{eq:mutipole-phase-profile} to the
SLM. The solid lines represent the expected core profile as given
by the LC. Blue and red lines surround, respectively, regions of negative
and positive LC. The solid lines have only the maximum radius as an
adjustable parameter, and since $\alpha=40$ for all $j$, the same
value was used for all curves. Figs. \ref{fig:multipoles in z0} (b,
e, h) show the LC, $\partial\Phi_{T}/\partial\phi$, as determined
from the experimental data \cite{Characterization_of_TC_and_OAM_in_SOV},
and Figs. \ref{fig:multipoles in z0} (c, f, i) exhibits the expected
LC according to Eq. \eqref{eq:mutipole-phase-profile}. A good agreement
is found between the theoretically expected results and the experimental
findings. The disagreement exists only at the darkest regions near
the profile center, where we were not able to properly measure the
phase. A technical aspect which is worth noticing is that the intensity
profile of multipoles is very sensitive to the spatial filter iris
transverse position, and misalignments makes the lobes profile nonsymmetrical.

\begin{figure}[h]
\begin{centering}
\includegraphics[width=8cm]{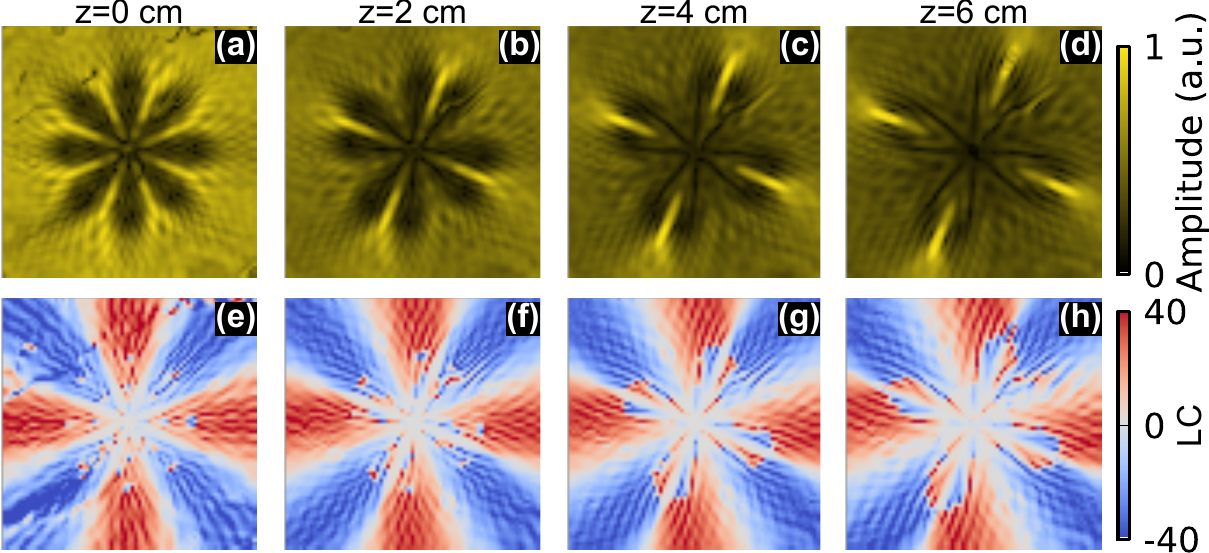}
\par\end{centering}

\protect\caption{Propagation of a vortex octupole $\left(j=4\right)$ in free-space.
In the top row (a-d) it is displayed the amplitude profile at increasing
propagation distance from the SLM image plane $(z=0\mbox{ cm})$,
while in the bottom row (e-h) is shown the experimental LC. Notice
that the LC remains stable under propagation.\label{fig:Multipole z evolution}}
\end{figure}

An important property of TC multipoles is that their LC is stable
under propagation, as can be seen by varying the position of the CCD
along the $z$ axis. Experimental amplitude and LC at different values
of $z$ are shown in Fig. \ref{fig:Multipole z evolution} for $j=4$.
It may be observed in the amplitude profiles, Figs. \ref{fig:Multipole z evolution}
(a-d), that pairs of amplitude lobes with opposed LC signs annihilate
under propagation. The resulting bright spots are located at zero
LC regions. Creation and annihilation of oppositely charged OV pairs
under propagation are well established in literature \cite{berrydislocations1974,Indebetouw1993,rouxdynamical1995,Swartzlandercomposite2003},
but to our knowledge the previous descriptions were always associated
with, respectively, creation and destruction of TC. In the case we
describe here, the beam's topological structure is preserved under
propagation, as can be seen from the LC in Fig. \ref{fig:Multipole z evolution}
(e-h). For negative $z$, the amplitude lobes rotate in the opposite
direction to that shown in Figs. \ref{fig:Multipole z evolution}
(a-d).

Another important property of the TP, Eq. \eqref{eq:Topological potential},
is that it satisfies a Gauss law inside a region $D$ for the total
enclosed TC, 
\begin{align}
\oint_{\partial D}d\vec{l}\cdot\vec{\nabla}V\left(\mathbf{r}\right) & =2\pi i\int_{D}da'\rho\left(\mathbf{r}'\right).\label{eq:Gauss_Law}
\end{align}

Equation (\ref{eq:Gauss_Law}) is a generalization of the usual expression
for the winding number over the phase profile, where one would have
$Q_{T}$ instead of $\int_{D}da'\rho\left(\mathbf{r}'\right)$ as
the total TC enclosed by $\partial D$. However, since Eq. \eqref{eq:Gauss_Law}
was obtained from a continuum generalization, care must be taken in
cases of continuous $\rho$. The discrete nature of TC makes $\rho$
an effective TC density.

To exemplify the meaning of $\rho$ in the continuous case we produced
a radial distribution for simplicity. We consider that $\rho=\rho_{0}r^{n}$,
where $\rho_{0}$ is constant, over a circle of radius $a$ and the
total TC distributed is $Q_{T}$. Since now $\rho$ is distributed
along a large region, the approximation of $r\sim0$ is not valid,
and the amplitude profile is not described by Eq. \eqref{eq:vortex_shape}.
However, the phase profile from Eq. \eqref{eq:Topological potential}
is always valid and equal to 
\begin{equation}
\Phi_{T}=\begin{cases}
Q_{T}\left[\left(\frac{r}{a}\right)^{n+2}\left(\phi-\pi\right)+\pi\right] & ,\, r<a,\\
Q_{T}\phi & ,\, r\geq a,
\end{cases}\label{eq:Phase_radial_TC}
\end{equation}
where we substituted $\rho_{0}=Q_{T}\left(n+2\right)/(2\pi a^{n+2})$
and it is assumed that $n\geq-2$.

Equation (\ref{eq:Phase_radial_TC}) is a generalization which smoothly
connects usual OV $(n=-2)$ to helico-conical beams, or optical twisters
\cite{Gluckstad_helico_conical2005,Gluckstad_OpticalTwister2011}
in which $n=-1$. Optical twisters are interesting because they carry
angular momentum and also have a higher photon density than the usual
Laguerre-Gauss or Bessel beams \cite{Gluckstad_OpticalTwister2011}.
Therefore they are of interest for manipulating particles \cite{Gluckstad_OpticalTwister2011}
and may also be of interest to nonlinear optics of OAM carrying beams
\cite{desyatnikovazimuthons:2005}. To our knowledge, other values
of $n$ were never previously reported in the literature.

\begin{figure}
\begin{centering}
\includegraphics[width=8.5cm]{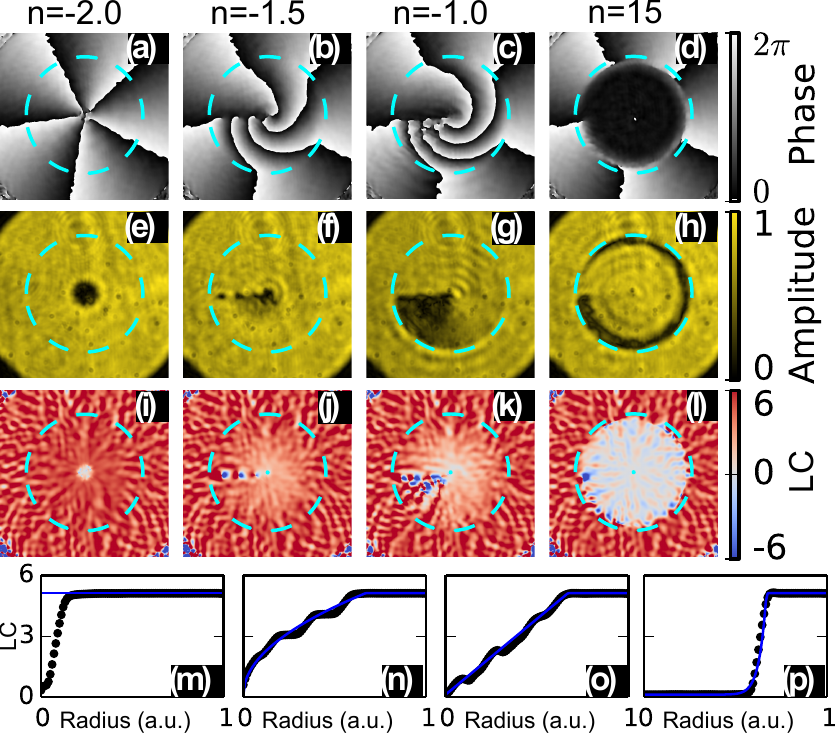}
\par\end{centering}

\protect\caption{Data for 2D radial distribution, Eq. \eqref{eq:Phase_radial_TC},
with $Q_{T}=5$, fixed $a$ and varying $n$ (columns) at $z=0$ cm.
$a$ is graphically represented by cyan dashed lines. The determined
LC (black dots) at (m-p) corresponds to the azimuthally averaged LC
at a given radial distance from the center of the circle as a function
of the radial distance. The blue solid lines in (m-p) correspond to
the values expected from the applied phase mask. \label{fig:2D_LWD_distribution}}
\end{figure}

We produced beams with the phase profile given by Eq. \eqref{eq:Phase_radial_TC},
with $Q_{T}=5$, fixed $a$ and varying $n$, and the results are
shown in Fig. \ref{fig:2D_LWD_distribution}. In the phase profiles,
Fig. \ref{fig:2D_LWD_distribution}(a-d), it can be seen that larger
$n$ increase the phase twisting at $r<a$ and reallocates the TC
towards the border along $\phi=\pi$. The TC displacement can be seen
also in the zeros of the amplitude profiles, as shown in Fig. \ref{fig:2D_LWD_distribution}(e-h).
The LC profiles, Fig. \ref{fig:2D_LWD_distribution}(i-l), shows that
larger $n$ values decrease the LC near the center of the circle.
In Fig. \ref{fig:2D_LWD_distribution}(m-p) we show the mean LC as
a function of the radial distance to the center of the circle. The
values obtained (black dots) agree with the theoretically expected
from the phase profile (solid blue) via Eq. \eqref{eq:Phase_radial_TC}.
The LC reduction in the center can be understood by considering that
larger $n$ push $\rho$ to the boundaries of the circle $a$. Therefore
the LC profile is directly associated with the $\rho$ distribution.
Also, since the LC is proportional to the local classical OAM of the
beam \cite{Characterization_of_TC_and_OAM_in_SOV}, this indicates
that the classical OAM profile depends similarly on $\rho$. 

In summary, we introduced in this work the concept of the Topological
Potential (TP), Eq. \eqref{eq:Topological potential}, by performing
conformal transformations over screw dislocations \cite{berrydislocations1974}.
The identification of the TP paves the way for further understanding
and tailoring of OV because it creates a bridge between OV and 2d
electrostatics. For applications where the shape of an OV is relevant,
as in optical tweezers \cite{DholakiaShapingOpticalManipulation2011},
laser traps \cite{Pruvost2010} or atom guides \cite{DholakiaAtomGuiding2006},
the TP might be used to design OVs for specific applications \cite{Shaping_Optical_Beams}.
Shaped OV may also allow selective excitation of plasmonic modes \cite{brasselettopological2013}.
However, while the present work can be directly used for OV at the
focus, further development is necessary to understand the effects
of propagation and address issues as the stability of OV \cite{AbramochkinSpiralLightBeams2004}.

Another important point is that the discussed examples obtained from
the TP might be useful in some applications. For instance, the intensity
profile instability of TC multipoles may be used to determine the
position of an extended object image plane, and in aligning spatial
filters. Another possibility is that, since multipoles form a complete
(Fourier) basis of orthogonal modes on the azimuthal phase, they may
be suited for applications in quantum communications \cite{WalbornAlignment-FreeOAMQuantumComm2012}.
In telecommunications, they are an alternative to Laguerre-Gauss beams
for data multiplexing \cite{WillnerTerabitFreeSpace2012} that can
be more stable to turbulence \cite{WillnerTerabitFreeSpace2012,WangAtmosphericTurbulence},
since the topological information is distributed over the beam profile.
The 2D TC distributions might be used to locally adjust the LC, and
consequently the local OAM, of a light beam by calculating Eq. \eqref{eq:Topological potential}
analytically or numerically. This is specially interesting for optical
tweezers, because then it becomes possible to locally adjust the light
induced torques. Therefore, one may in principle control trapped particles
in 2D by simply adjusting the TC distribution.

We acknowledge the financial support from the Brazilian agencies CNPq
(INCT-Fotônica) and FACEPE. A. M. A. also acknowledge Tiago T. Saraiva
for the helpful discussions.

\end{document}